# Experimental Investigation of Crack Jumps during Initiation and Growth in IN718

A. Pokharel*, J. Lindsay*, S. Papanikolaou*, T. Musho*+
* Department of Mechanical & Aerospace Engineering, West Virginia University, 395 Evansdale Drive, Morgantown, WV, 26506-6070
+ Corresponding Author – tdmusho@mail.wvu.edu

## Abstract:

The following study investigates the statistical nature of crack jumps during fatigue of Inconel 718. In-situ measurement in atmospheric air of the crack length at several loading and temperature conditions was conducted using a direct current potential drop (DCPD) method. Both annealed and heat-treated Inconel 718 samples were investigated. For a single sample, the normalized change of crack length was confirmed to not be a random process. This finding is significant in highlighting the time history features, which could be used in training machine learning models for the fundamental understanding of oxidation and the prediction of crack initiation and growth.

Keywords: Crack Propagation; DCPD; Inconel 718; Fatigue

## Introduction:

With the recent move to reduce carbon emissions across many counties and the constant goal of reducing operation costs, gas turbine efficiency can achieve both goals and has been increasing steadily over the past 50 years [1]. One of the methods that have been employed to increase the gas turbine efficiency is to have turbines operate at higher temperatures [2]. However, higher temperatures require new materials and alloys, as well as a better understanding of the basic physics that may drive materials informatics [3]. Regarding mechanical response, a better understanding of this physics in various materials has coincided with the development of new predictive models for failure mechanisms. A common superalloy used in the disk of turbines is Inconel 718 (IN718), a nickel-based alloy that is designed to withstand extreme temperatures under high creep load conditions. The main constitutive elements in this type of superalloys are nickel, cobalt, and iron with a precisely prescribed precipitate-hardened microstructure that provides high strength, high operating temperatures, as well as creep and oxidation resistance. It is a combination of constitutive elements and microstructure but more so the microstructure that impedes and governs the crack initiation and growth in this material. However, structural health monitoring has been typically limited in qualitative assessments of isolated microstructural observations. Beyond the study of averages, a fundamental understanding of crack initiation and growth, as well as structural health assessment, requires new approaches that focus on connecting experiments and advanced simulations in the statistical frontier. In this study, the objective is to monitor the dynamics of crack initiation and growth in an IN718 sample under monotonic low-cycle fatigue loading at various operating temperature and use it towards a history-informed, as well as statistically-informed assessment of structural health. More specifically, the focus of this work is to capture the statistical variation of the magnitude of crack length growth versus cycle number during initiation and growth stages.

In this study, IN718 has been selected as the material of focus due to its polycrystalline nature and wide application in contemporary turbines [4]. IN718 has a wide range of operating temperatures from -423°F to 1300°F while maintaining strength, good mechanical properties of weldability, and resistance to cracks commonly caused by welding [5]. The key to IN718's performance is the compositional complexity of 15+ constitutive elements including titanium, cobalt, niobium, etc., as well as a production processing stage that optimizes the precipitate hardening capacity of the alloy [5]. The important aspect of IN718 is the strengthening phases of the microstructure. The primary strengthening phase is the γ" phase, which is a metastable phase, with the δ phase being the thermodynamically favorable phase. Since the γ" phase is not thermodynamically favorable, the over-aging cooling rates must be precisely controlled. During this non-equilibrium cooling process, the γ" phase partially orders in disk-shaped precipitates that are coherent with the γ phase. The origin of the ordering lies in the emergence of coherent strains from the lattice distortion of the precipitate formation [6]. In addition to γ", there are also secondary phases γ' and γ present within IN718. The γ' phase also plays an important role as a strengthening phase in IN718, but to a lesser degree than γ". The γ" phase is on the order of four times larger than the γ' phases [7]. The γ' is often found as a fine dispersion of spherical particles, which are also coherent with the γ phase. It is interesting to point out that the coherency of these precipitate phases should potentially be relevant in the origin and magnitude of crack length jumps as fatigue progresses.

The main focus being the experimental capture of the crack during initiation and growth. For this purpose, there are several experimental methods: The simplest method involves the use of a high-resolution optical microscope, which is typically limited to distinguishing up to submicron features [8]. Another common method is the use of an optical microscope in conjunction with another measurement method such as a direct-current potential drop (DCPD) method [9]. Other approaches include *in-situ* SEM of crack growth; however, this approach requires special equipment that integrates a field emission gun with a tensile testing apparatus [10]. Digital Image Correlation (DIC) measurements can also take *in-situ* measurements of crack growth [11, 12] along with the added benefit of being able to measure the strain field of the specimen [13]. The measurement method selected for this study is the ASTM recommended direct current potential drop (DCPD) method [14] to measure *in-situ* short crack growth in low cycle fatigue crack-growth.

# Experimental Method

The experimental method employed for this study is based on the combination of the DCPD method and fatigue loading of a statistically significant (n≥17) number of IN718 samples in a hydraulic MTS load frame. We employ the ASTM compact specimen standard testing procedure with a custom specimen geometry. The test specimens were not run to failure but were stopped prior to the ultimate fracture crack growth point. Samples were initiated with an electro-discharge machined initial crack in the absence of a crack initiation procedure to capture the crack initiation event and the later stages of crack propagation.

**Test Specimen Design**

Following the ASTM compact specimen [14] design specification, a custom test specimen was designed based on a plane strain measurement using IN718 material. The design is a 76.2mm x 36.63mm simple rectangle of plate Inconel 718 with two holes for mounting the specimen. A schematic of the design is illustrated in Fig. 1A. Along with the two mounting holes, a small 10mm slit was cut using electrical discharge machining (EDM) with a width of 0.16mm. The EDM cut serves the purpose of acting as an initial crack in the material. The tip of this initial crack is rounded with a radius equivalent to the EDM wire or 0.08mm. In providing electrical continuity without introducing secondary materials nickel-chromium wires with a diameter of 0.4mm were spot-welded to the samples. Care was taken to spot-weld the wires in the same position on either side of the EDM. These wires closest to the EDM cut serve as the voltage probes. A larger gauge nickel-chromium wire with a diameter of 1.0mm was spot welded across the full specimen as current leads 15mm above and below the EDM cut. This wire placement provides a uniform potential across the specimen. This is analogous to the original DCPD experiment setup by H. Johnson who used copper clamps [15]. The current was held constant with a DC current power supply at 10A.

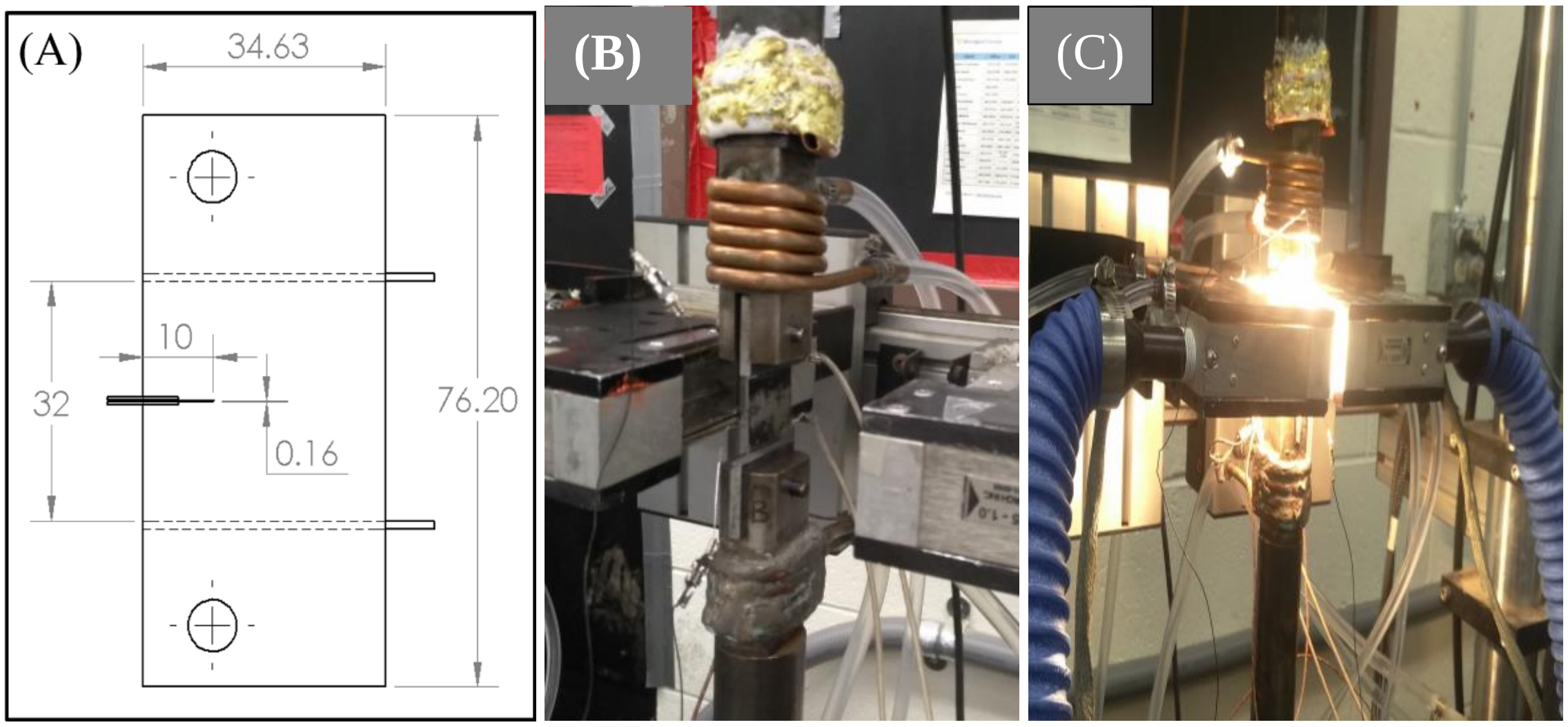

**Fig. 1** – *(A) Schematic of plane strain Fatigue Specimen which shows the EDM cut thickness and length, two pin holes for loading, and the location of the wires (units are in mm). (B) Setup of Fatigue Test showing the two white wires in the back which provide the 10 amps and the two smaller wires that probe the votage. (C) Setup of Fatigue Test showing the specimen being heated by two incandescent lamps located on either side of specimen.*

**Machine Setup**

The fatigue crack-growth testing is done on an MTS 810 hydraulic load frame. A custom fixture was machined to hold the specimen and allow both tension and compression of the sample, see Fig. 1B and Fig. 1C, to mount the specimen into the testing machine. The custom fixture was cooled at the top and bottom to avoid the thermal drift of the load cell. To ensure that the sample was positioned vertically, a steel spacer was used. Rolled pins were used to connect the spacer with the specimen. Thermocouples were mounted to the sample and the lamps were controlled at a set point temperature using a PLD controlled. The heating lamps that contained incandescent bulbs were cooled using a combination of air and water. Data from the experimental devices were collected using the MTS Flextest 40 digital controller, with the sampling rate set to 512Hz. It should be noted that this is a high sampling rate for DCPD, which was required to analyzed features in the electrical measurement. The potential was measured using the strain gauge amplifier built into the Flextest hardware. Data was exported as CSV files with a number of cycles, force, and voltage.

**Analytical and FEA Comparison**

To utilize the voltages taken from the DCPD measurement it is necessary to relate the voltage to a crack length. The relationship between the voltage and crack length is known as a crack length calibration curve. This relationship is a function of not only the geometry of the samples but also the placement of both the current and voltage probes on the specimen. In this study, two methods are employed and compared to derive the calibration curve. The first method is to follow the analytical calibration curve derived by H. Johnson shown in Equation 1. The second method is to use finite element analysis (FEA) to conduct an electrostatic simulation to predict the potential at the probes with varying crack lengths. The modern approach is to use FEA to derive the calibration curve as it improves the sensitivity at longer crack lengths [15, 16, 17]. Equation 2 is the derived polynomial that describes calibration curves based on the FEA results.

$$\frac{a}{W} = \frac{2}{\pi}\cos^{-1}\left[\frac{\cosh\left(\frac{\pi y}{2W}\right)}{\cosh\left(\frac{V}{V_0} cosh^{-1}\left\{\frac{cosh\left(\frac{\pi y}{2W}\right)}{cos\left(\frac{\pi a_0}{2W}\right)}\right\}\right)}\right] \tag{1}$$

$$\frac{a}{a_0} = 0.0102\left(\frac{V}{V_0}\right)^3 - 0.1787\left(\frac{V}{V_0}\right)^2 + 1.1917\left(\frac{V}{V_0}\right) - 0.02 \tag{2}$$

In Equation 1 and 2, $a_0$ is the initial crack length (10mm) created by the EDM cut, a is the total crack length, W is the specimen width, y is the plate thickness, $V_0$ is the initial voltage before crack propagation, and V is the measured voltage throughout the experiment. The result of both of these calibration curves is plotted in Fig. 2. There is high accuracy at short crack lengths (the length scale that this study covers) and good agreement between the analytical and FEA calibration curves. It has been elaborated in other studies [16] that FEA calibration has less error than the analytical calibration curve at longer crack lengths. For the remainder of this study, the FEA-derived calibration curve will be used.

Because the stress intensity factor is also a function of the geometry it is necessary to account for the sample geometry when predicting the stress intensity factor. There are several closed-formed expressions that use empirical geometry correction factors. In this study, an empirical expression for an edge crack under uniaxial stress [18] was considered. The corresponding stress intensity was calculated using the following expression,

$$K_I = \sigma\sqrt{\pi a} * \left[1.122 - 0.231\left(\frac{a}{w}\right) + 10.55\left(\frac{a}{w}\right)^2 - 21.71\left(\frac{a}{w}\right)^3 + 30.382\left(\frac{a}{w}\right)^4\right], \quad (3)$$

where σ is the uniform stress state and the factor rightmost polynomial is a geometrical factor.

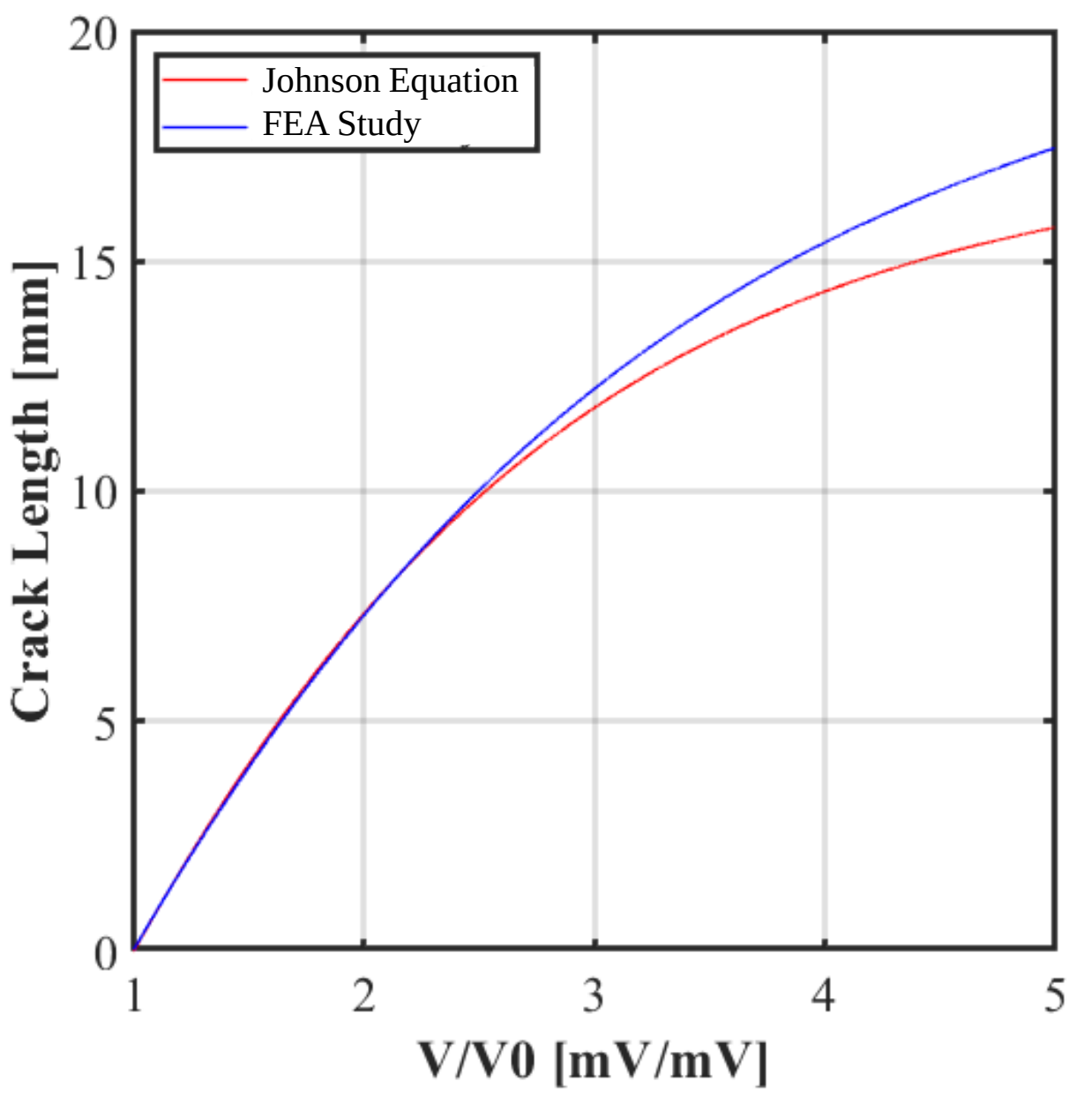

**Fig. 2** – Plot of the Johnson analytical DCPD calibration curve and the FEA derived curve. At short crack lengths the curve are in good agreement.

**Material Properties**

Two heat treatment conditions of IN718 were analyzed during this study. Those include the as-received heat treatment condition also confirmed to be the annealed condition. The second condition was the fully heat-treated condition, which satisfies the ASTM [19] condition for IN178. To determine the mechanical properties of the samples, prepared from two heat treatment conditions, test specimens were cut out and tested using a load frame according to ASTM tensile standard for ductile metals [20]. The resulting stress-strain curve can be found in Fig. S1 in the Appendix. The annealed condition provided an ultimate tensile strength of 770 MPa at 21% elongation and yield strength of 430 MPa with a modulus of elasticity of 185.4 GPa. Hardness testing was also conducted with a hardness testing machine which gave a Rockwell-C hardness of 21.1. It should be noted, that sheet stock of IN718 is typically provided from the manufacture in the annealed conduction and requires subsequent heat treatment.

The second heat treatment condition was obtained using a programmed heat-treated procedure with nitrogen quenching. The heat treatment procedure followed the ASTM standard for IN718. The first step of the heat treatment involved holding at 1775 $^{0}$F for an hour and cooled below 150 $^{0}$F in the nitrogen environment. Without opening the furnace, the samples were then held at 1350 $^{0}$F for 8 hours. This process was followed by the cooling of samples to 1150 $^{0}$F at the rate of 100 $^{0}$F per hour. After holding at 1150 $^{0}$F, until the total age cycle was 18 hours, the samples were cooled below 150 $^{0}$F using nitrogen. The samples thus prepared showed an enhanced mechanical property. The Rockwell-C hardness was determined to be 41. Also, shown in Fig. S1 in the appendix, the ultimate tensile strength was found to be 1330 MPa.

**Test Parameters**

The specimens were placed in the hydraulic testing machine using pin joints as shown in Fig. 1B and Fig. 1C. Because of the thin design of the specimen, small still spacers were used to align the sample and mitigate tear-out. Because the study is interested in crack initiation, the specimens did not undergo a crack initiation procedure. In other studies, this is typically done to speed up the initiation of the crack by running at higher stress intensity without a hold cycle to initiate the crack. In this study after the sample was loaded in the machine, data acquisition began with a loading cycle that was used were made up of two parts, 10 fast oscillations for 30 seconds at a given peak load and that cycled with a ratio between the max and min of R=0.05. Following the oscillation, a 100-second hold with loading at the peak load was carried out. For this study, the peak loads were selected as 1600N, 1700N, and 1800N. More emphasis is given to the three highest loadings.  The representative loading cycles as a function of time can be seen in Fig. 3. It should be pointed out that the target loading ratio (R=0.05) was specified so that there is a small difference between the true min-max range for the oscillations, as shown in Table 1 below. The actual ratio for all the cases was R=15%. This was a limitation of the control software and inertia of the machine to precisely achieve both the theoretical min and max values. Also, the experiment was specifically conducted at three different temperature: room temperature (RT), 185$^{o}$C, and 285$^{o}$C.

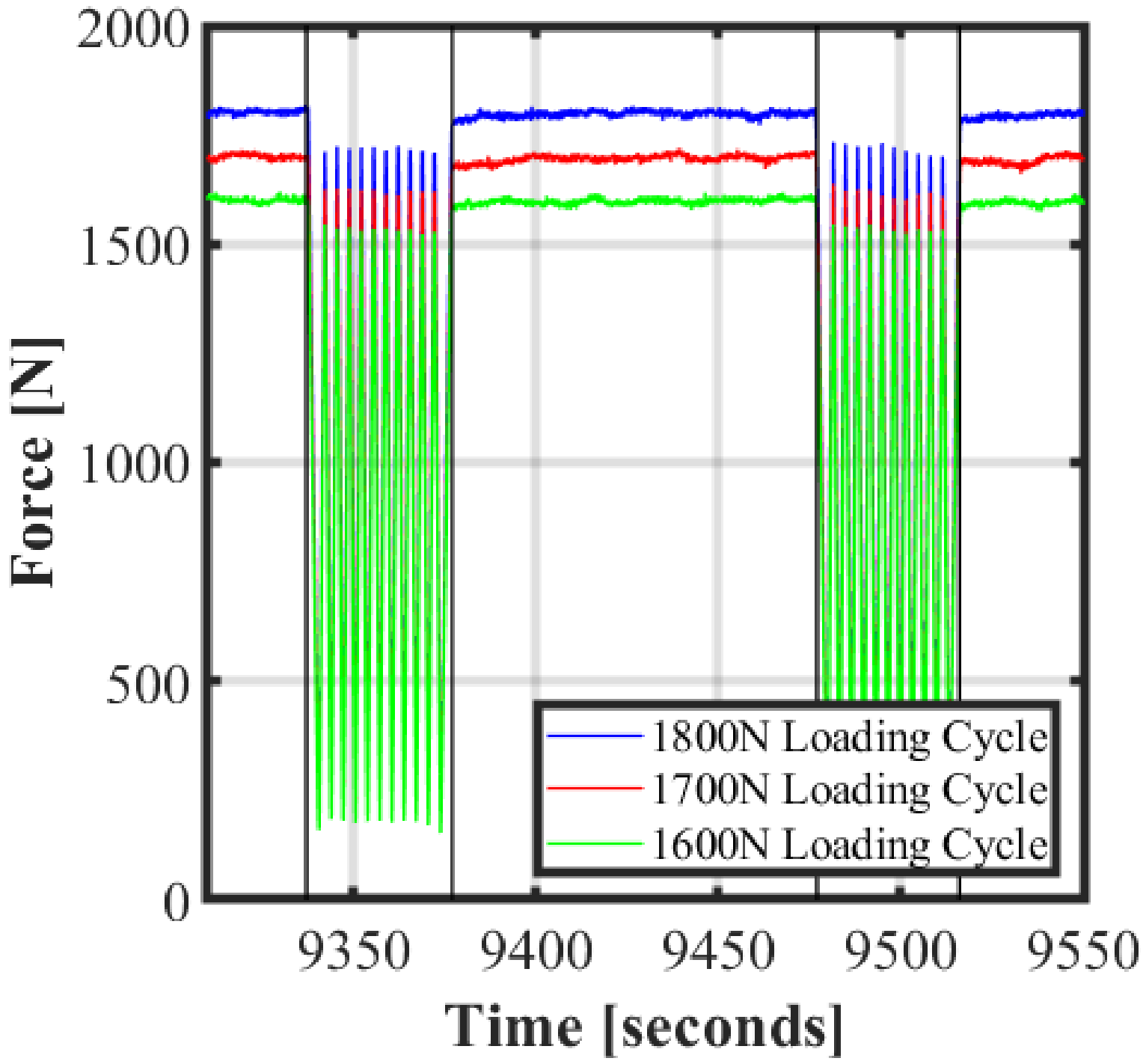

**Fig. 3** – Plot of the transient loading cycles for three loading cases. One cycle consisted of 10 oscillations between $F_{max}$ and $R*F_{max}$ over a thirty second period, followed by a 100 second hold at $F_{max}$.

**Data Processing**

Data acquisition was conducted at a high sampling rate of 512Hz (~2ms/sample) to capture the crack jumps that occur at small length scales and small durations. The data collected was post-processed using Python. To correlate the measured potentials to crack lengths the aforementioned calibration curves were employed along with the stress intensity factor relationship, see the previous section for details. For the 1800N peak load case n=17, for 1700N n=18 and for 1600N n= 22. The data was post-processed by examining the 100-second hold cycles and computing the difference between the crack length over subsequent hold cycles. The difference in crack length between subsequent hold cycles is considered a jump off the crack. For this study, the exact time of occurrence during the hold cycle was not of interest but rather just the quantity and statistics of jumps between subsequent steps. Information about the exact time to the nearest +/-1ms during the hold cycle is available due to the high sampling rate.

| | *Theo. Osc. Min (N)* | *Theo. Osc. Max (N)* | *Theo. Osc R (N/N)* | *Act. Osc Min (N)* | *Act. Osc. Max (N)* | *Act. Osc. R(N/N)* |
|---|---|---|---|---|---|---|
| *1600N* | 85 | 1600 | 0.05 | 210 | 1430 | 0.15 |
| *1700N* | 90 | 1700 | 0.05 | 230 | 1570 | 0.15 |
| *1800N* | 95 | 1800 | 0.05 | 250 | 1630 | 0.15 |

**Table 1** – Summary of the theoretical and actual min and max forces specified for fatigue cycles. The actual ratio is based on the DAQ from the load frame. The theoretical values are the input control values to the load frame. Note, a R=0.05 was specified by an R=0.15 was realized.

An aspect inherent to this DCPC experimental measurement method is the incorporation of electrical noise in the measurement. This electrical noise is a product of the environment and was minimized by using condition DC power supplies. It is important to point out that the electrical noise in the experiment is random noise. By plotting a histogram of the noise and evaluating the p-value, see Fig. S2 in the appendix, it was found that the electrical noise is indeed random and follows a normal distribution. This is an important claim because the non-random noise in the DCPC measurement signal is, therefore, a result of some attribute in the material, more specifically, the crack.

# Results and Discussion:

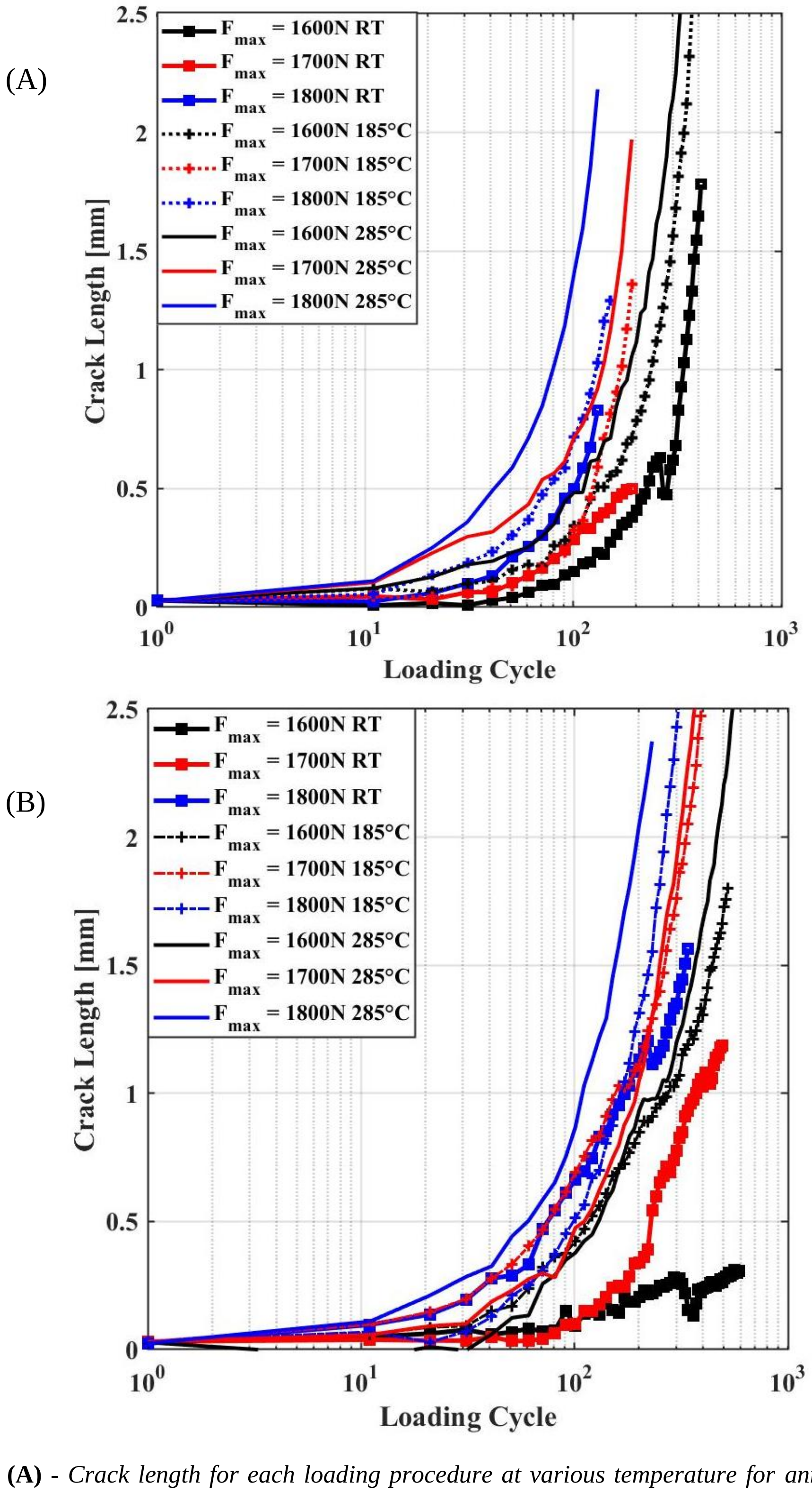

**Fig. 4 (A)** - *Crack length for each loading procedure at various temperature for annealed IN718 samples. Lines are associated with different peak loads or initial stress intensities.* **Fig. 4 (B)-** *Crack length for each loading procedure at various temperature for heat-treated IN718 samples. Note the annealed samples faster initiation and growth.*

Following the experimental procedure outlined above, the average crack length for each subsequent holding cycle was calculated. It should be clarified for the discussion that the terms crack and crack length refer to the crack originating at the end of a 10mm EDM cut. Fig. 4 is a semi-log plot that illustrates the crack length as a function of cycle number with various loading conditions and different operating temperatures. All tests were stopped prior to Region III. Fig. 5 illustrates crack initiation and growth from Region I through Region II. The values in the key of Fig. 5 correspond to the peak loading values, which can be related to the initial stress intensity. As expected, as the peak load increased the stress intensity increased and the rate of crack propagation also increased. It should be noted, in Fig. 4, that the samples were not crack initiated, and it takes an increasing amount of run time or cycles depending on the loading to initiate the crack. This was a translation of the curves in the x-axis of Fig. 5 and an associated delay in lift-off from the x-intercept. To provide some context to the wall time required for all the cases, the 1800N case for the annealed sample at 285$^{o}$C took approximately 6.5 hours to reach a crack. Further, the $\Delta a/\Delta N$ vs $\Delta K$ graph for the annealed sample is presented in Fig. S6 in the appendix.

In addition, crack length was found to be commensurate with the operating temperature. From Fig. 4, it can be seen, the crack length of a high-temperature specimen is longer as compared to the lower temperature specimen operating at the same loading condition and the same number of current hold cycles. Crack propagation was found to be increased with an increase in the temperature of the specimen. To illustrate: for a specimen at RT in Fig. 4 (A), it required 412 current hold cycle under 1600N loading condition to reach the crack length of ~1.8mm while it only took 350 current hold cycle for a specimen under the same loading condition to reach the crack length of ~2.8mm operating at a temperature of 285$^{o}$C.

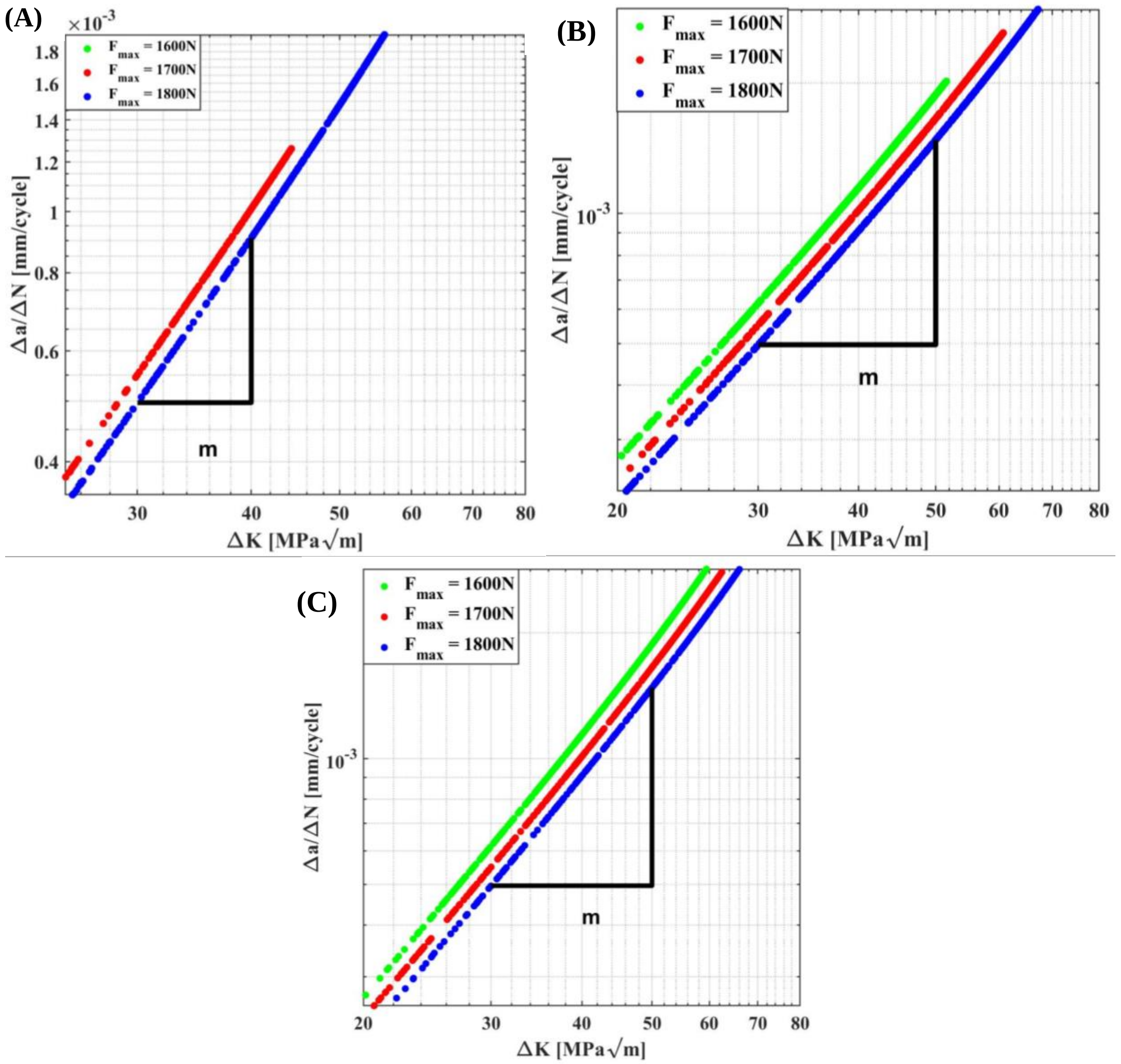

**Fig. 5(A)** - *Plot of Δa/ΔN versus the stress intensity factor difference for multiple loading cycles at RT for heat-treated IN718 samples* **Fig. 5(B)** - *Plot of Δa/ΔN versus the stress intensity factor difference for multiple loading cycles at 185°C for heat-treated IN718 samples.* **Fig. 5(C)** - *Plot of Δa/ΔN versus the stress intensity factor difference for multiple loading cycles at 285°C for heat-treated IN718 samples. The data from all experimental datasets demonstrate Paris' Law regime.*

**Crack Characteristics**

Taking the information from Fig. 5 and casting the data in the form of rate of crack length versus stress intensity, it is possible to determine which regime of fracture the crack was predominantly undergoing. All samples tested appear to spend the majority of the experiment time in Region II, the Paris Law region [21], while the test was stopped prior to Region III [22]. Fig. 5 illustrates the linear nature of the Paris Law for all samples [23]. The exponent of the Paris Law equation was calculated to be about 2.7 with the coefficient as $10^{-5}$, this agrees with what has been seen in literature [24]. Less time was spent in Region I due to the design of the specimen and the position of the load line relative to the EDM cut. Comparing across Fig. 5 as the temperature increases the fatigue resistance decreases. This is a well-characterized phenomenon in air environments. While it is beyond the scope of this study it is well accepted that oxidation plays a critical role in high-temperature fatigue. That being said, many of these features in the DCPD measure could be attributed to the role of oxidation in high-temperature fatigue tests.

**SEM Imaging and Analysis**

SEM images were taken of the samples after the fatigue tests were completed. Fig. 6 is a collection of images from a single specimen. Fig. 6A illustrates that the crack had initiated at the back of the notch that was created by the EDM cut. This was the expected and desirable position of the crack and confirms that the samples are being loaded symmetrically. Fig. 6B is an image of the crack at the mid-section of the crack. Looking at the crack surface striations or beach lines are present, which is synonymous with fatigue [25]. Further examination of the striations reveals that the 1μm+/-0.05μm, which was on the order of the $\Delta a$ measured using the DCPC reported in Fig. 4 and 5. Fig. 6B also depicts that the fracture surface was not smooth, which is representative of a more ductile fracture as opposed to a brittle fracture. In Fig. 6 D-F for the heat-treated sample, many of these features change as the sample becomes more brittle with heat treatment.

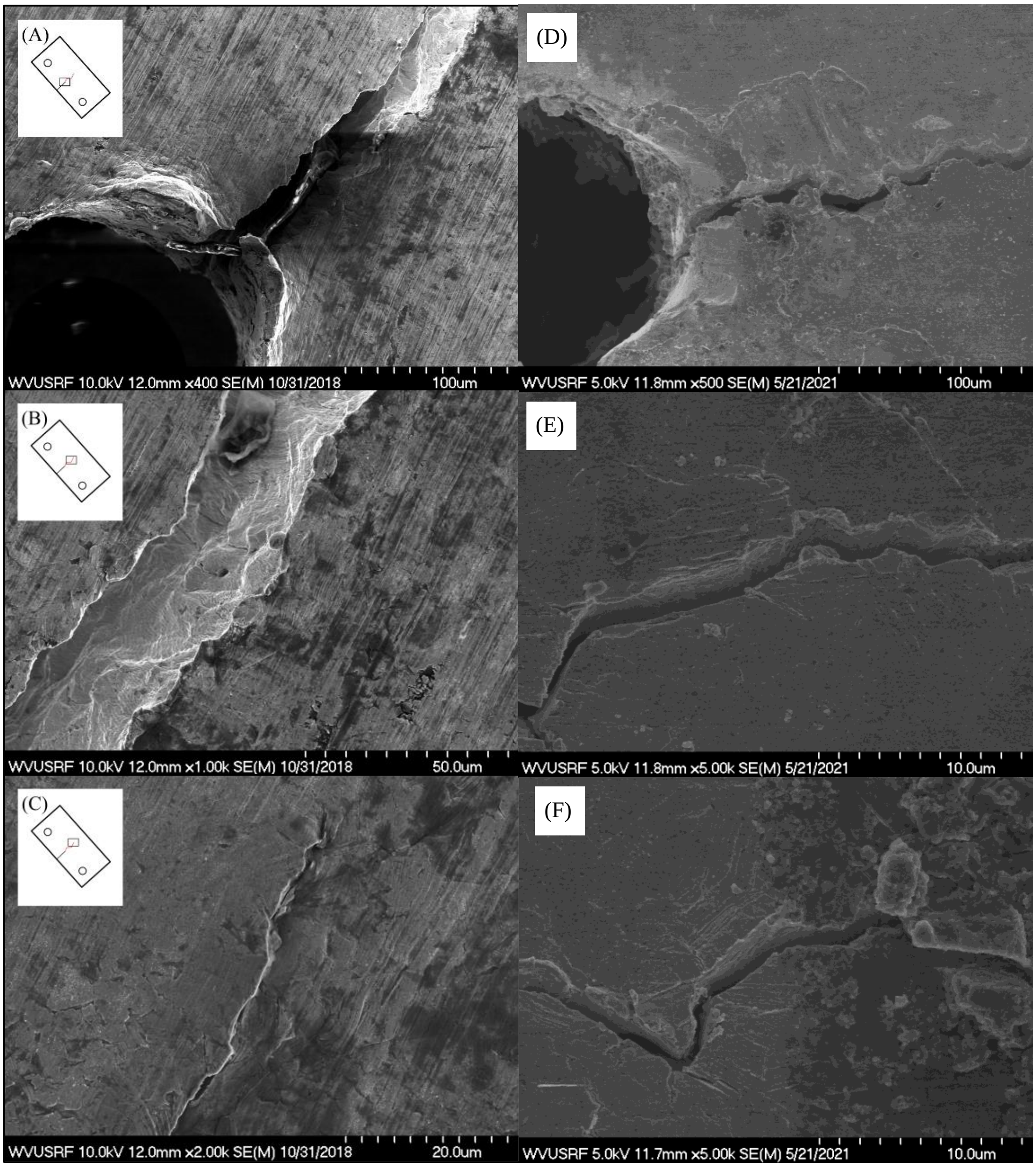

**Fig. 4** - *Crack initiation point icrack initiation at the back of the EDM cut for annealed (**A**) and heat-treated (**D**) sample.* **Fig. 6(B)** - *Zoomed in image along the crack path at mid-length with visible beach marks with frequency of 1µm for annealed (**B**) and heat-treated (**E**) sample.* **Fig. 6(C)** - *Termination zone of crack which shows evidence of slip bands at the very tip for annealed (**C**) and heat-treated (**F**) sample.*

Fig. 6C was an SEM image at the crack termination prior to Region III. In Fig. 6C persistent slip bands (PSB) are evident near the crack tip, which is directed away from the crack tip. Typically, during crack initiation, the crack will run along with the plane of the PSB, and during crack propagation the PSBs position ahead of the crack in the plastic zone. It can be seen from Fig. 6C that the PSB is ahead of the crack tip and therefore it can be visually confirmed by the relative orientation of the PSB that the crack was beyond the initiation phase when the loading is stopped. More interesting was the extent of the plastic zone on either side of the crack extends approximately 8um to either side. Further investigation of the band's width reveals that they are on the order of 0.5μm, which is similar in magnitude to the beach marks found in Fig. 6B. It is not obvious how these PSB influence the electric field around the tip, which is an area of future investigation.

**Noise Qualification and Quantification**

A critical aspect of this study was to qualify and quantify the noise of the crack during fatigue. As mentioned previously, random electrical noise was present from the DCPD method and it is the non-random noise that is of interest. It was expected for a truly random event such as electrical fluctuations from the DCPD method that the distribution should follow a normal or Gaussian distribution that is centered about a zero mean provided the systematic bias has been removed through calibration. This was confirmed by running a zero-load test and creating Fig. S2 which illustrates a standard Gaussian distribution. With this information, the standard deviation of the noise was used to calculate upper and lower bounds for the $\Delta a/\Delta N$ curves shown in Fig. 7. These error bars showed that below a certain loading threshold, the signal to noise ratio becomes too much and starts to bury the signal change from the crack growth.

To account for the increase in stress intensity as the crack length increases it was necessary to normalize the change in crack length by the stress intensity for each case. Fig. 7 is a plot of the resulting normalized crack length as both a function of the cycle and in a histogram to show the spread of the crack growth. Fig. 7A shows the change in crack length versus loading cycle for a single run, while Fig. 7B shows the average across all the samples that were run with a max loading of 1800N. The rightmost graphs of Fig. 7A and Fig. 7B is a plot of the normalized crack length growth as a function of the cycle with shaded regions to show the first standard deviation of the electrical noise. A reference Gaussian distribution was added to each histogram using the mean and standard deviation of the righthand plots. This was to show the agreement of the histograms to a Gaussian distribution while also showing the change in the magnitude of the mean with different loading cases.

Fig. 7A shows that for a single case, the crack growth seems to follow a different distribution than a Gaussian distribution. To validate this observation normality test was conducted for the experimental data. The null and alternative hypothesis was developed; the former corresponds to the decision that the experimental data follows the normal distribution while the later discard it. The significance level, normally known as alpha, was set to 0.05, and the p-value for the experimental data was calculated. Based on the comparison of p-value with alpha, the decision to choose between null and alternative hypothesis are made. The null hypothesis is validated if the p-value is greater than alpha else the alternative hypothesis is accepted. Our calculation for the experimental data, shown in Fig. 7A and Fig. S3-S5, showed a significantly lower p-value as compared to alpha. Hence, the null hypothesis was discarded, and an alternative hypothesis was chosen. While in Fig. 7B, the averaged crack growth values showed good agreement with a Gaussian distribution. This difference between single cases and the averages is due to transient independent crack growth events that are not shared by multiple tests. These independent events contain information about how the microstructure reacts to the crack growth that is typically lost when looking at the averaged trends of multiple sets of tests. Because these independent events correspond to reactions of the microstructure to the growth of the crack the time history is unique for each sample.

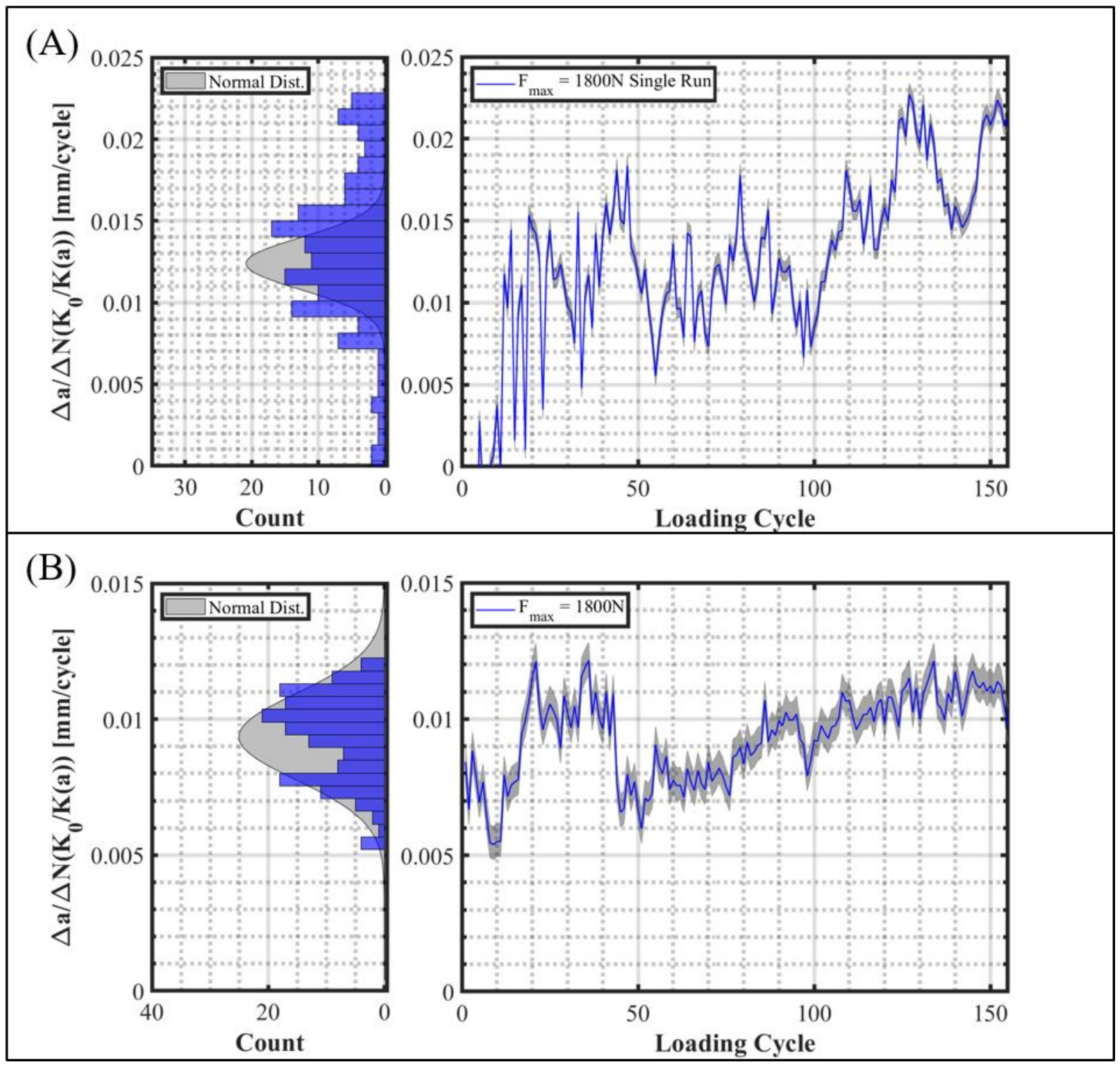

**Fig. 7(A)** - *$F_{max}$=1800N plot of crack jump distribution where the rightmost figures are a plot of the normalized crack jumps over subsequent hold cycles and the leftmost figures are a histogram of the crack jumps.* **Fig. 7(B)** - *$F_{max}$=1700N plot of crack jump distribution.*

Fig. 8 is a plot of the relative probability distribution ($v_i = c_i/N$, where $v_i$ is bin value, $c_i$ is the count, and $N$ is the total number of elements) with data taken from the leftmost subfigures of Fig. 8 in addition to additional loading cases. Fig. 8 further illustrates the trend that for increased loading there is increased magnitude in the crack jumps. Moreover, Fig. S2 provides not only a relative reference to other loadings but also to zero. Comparing the range to each of the other distributions, as the loading increases the range distribution decreases.

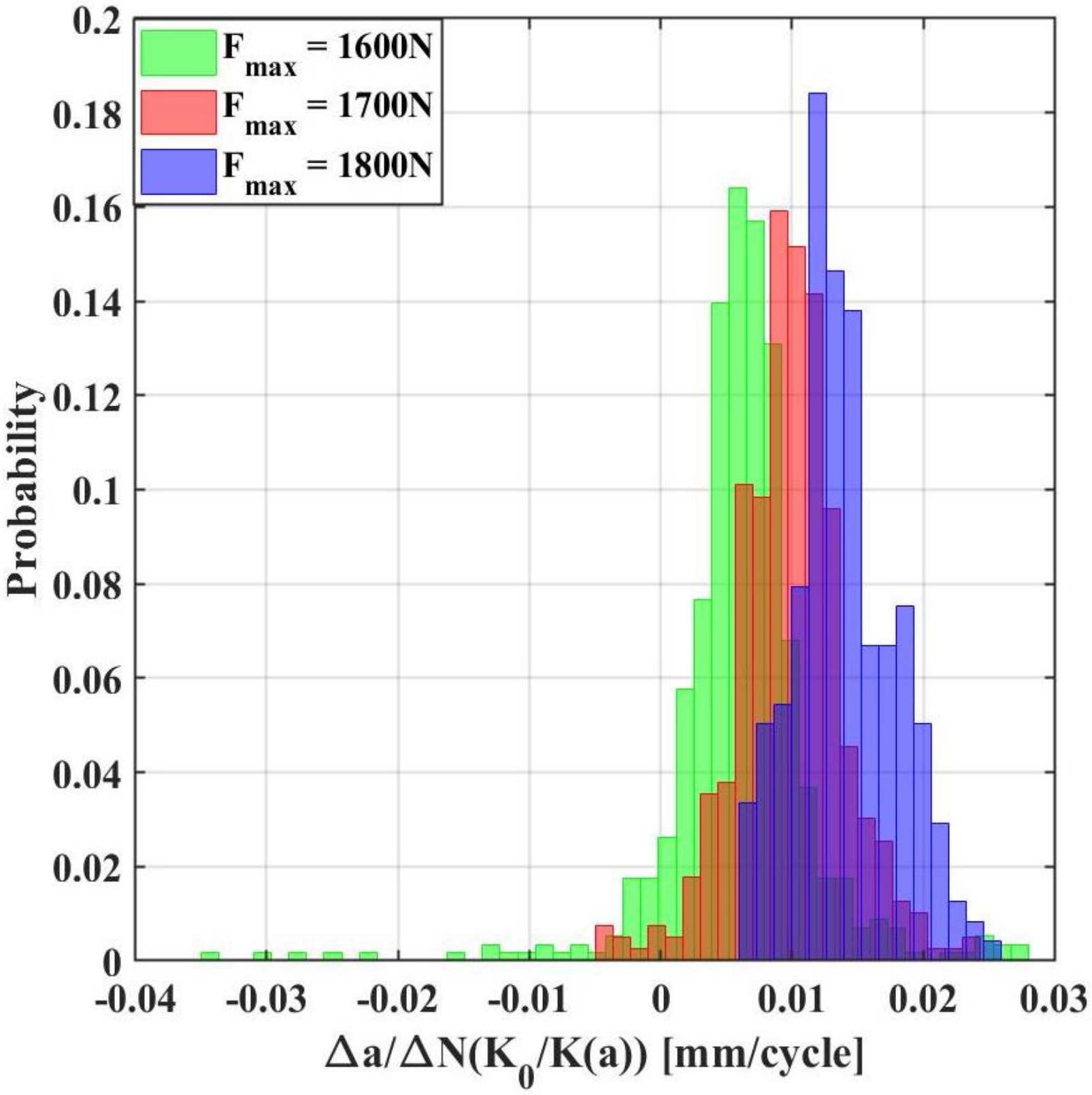

**Fig. 8** - *Relative probability distribution of crack jumps for loading cycles ranging from 1600N to 1800N. Increasing peak load results in increased magnitude in jumps and decrease in distribution range.*

Additionally, as the peak loading decreases below 1600N there is an appreciable number of negative length crack jumps. While these negative crack lengths can be reasoned by the crack closure this most likely not the case because the minimum loads are never zero and therefore the crack should not close. The most reasonable explanation of the negative values is that at lower loading the noise associated with the electrical measurement is dominating and the distribution is becoming Gaussian. Another way to explain this is in terms of the signal-to-noise ratio and the fact that the ratio is decreasing for decreasing load. This is an important finding in that it provides a discrete cut-off for the signal-to-noise ratio of the proposed DCPD method.

Capturing the statistical variation of the crack jumps for different loading cycles provides an argument for using DCPD data in a more complex simulation. More specifically the rightmost graphs of Fig. 7 show a large quantity of data down to the millisecond that is has captured important features of the underlying physics. As mentioned previously an important aspect of high-temperature fatigue is the role of oxidation. It is reasonable to assume that information about the oxide formation and its influence is captured in the DCPD. One such tool that is now being explored that could potentially deconvolute these features that are captured in the use of machine learning algorithms.

# Conclusion:

This study focused on conducting a DCPD measurement on IN718 samples at several peak loads at various operating temperatures for two heat treatments. This study confirms that there are unique features contained within individual time histories of DCPD measures. Often these features are lost when multiple DCPD measurements are taken and averaged. In this latter case, the DCPD measurement becomes a random process. Using the normality test it was found that the data for a single run experiment is not a random process. Hence, an event early on in the time history will influence an event later in the time history. With this finding, it can be concluded there the time history of individual DCPD measurements provided information such as how oxidation influences high-temperature fatigue. It is hypothesized that a machine learning model is a potential candidate for training on these DCPD features, relating them to the underlying physics.

# Acknowledgment:

The authors would like to recognize the funding support from the National Science Foundation Award #1709568. T.M. would like to acknowledge the experimental help of Lucas Ware and Liam Thomas. T.M. would also like to acknowledge the WVU Shared Research Facility technician Marcela Redigolo for taking SEM images.

# Competing Interest Statement:

None of the authors have a competing interest in the presented work.

# References:

[1] Kyprianidis K. *Future Aero Engine Designs: An Evolving Vision, Advances in Gas Turbine Technology*. IntechOpen. 2011.

[2] Farokhi S. *Aircraft Propulsion.* Chichester (UK): John Wiley & Sons; 2014.

[3] Downs J, Landis K. *Turbine Cooling Systems Design: Past, Present and Future*. ASME Turbo Expo 2009: Power for Land, Sea, and Air. 2009.

[4] Reed R. *Superalloys Applications.* The Minerals, Metals & Materials Society (TMS). February 2007.

[5] Special Metals, "INCONEL® Alloy 718," 2007. [Online]. Available: http://www.specialmetals.com/assets/smc/documents/alloys/inconel/inconel-alloy-718.pdf.

[6] Hong S, Chen W, Wang T. *A diffraction study of the γ "phase in INCONEL 718 superalloy*. Metallurgical and Materials Transactions A. 2001; 32: 1887-1901.

[7] Slama C, Abdellaoui M. *Structural characterization of the aged Inconel 718*. Journal of alloys and compounds. 2000; 306: 277-284.

[8] Davidson M. *MicroscopyU.* Nikon. [Online]. Available: https://www.microscopyu.com/microscopy-basics/resolution.

[9] Chen D, Gilbert C, Ritchie R. *In Situ Measurement of Fatigue Crack Growth Rates in a Silicon Carbide Ceramic at Elevated Temperatures Using a DC Potential System*. Journal of Testing and Evaluation. 2000; 28: 236-241.

[10] Zhang W, Liu Y. *Investigation of incremental fatigue crack growth mechanisms using in situ SEM testing. International Journal of Fatigue*. 2011; 42: 14-23.

[11] Roux-Langlois C, Gravouil A, Baietto M, Rethore J, Mathieu F, Hild F, et al. *DIC identification and X-FEM simulation of fatigue crack growth basedon the Williams' series*. International Journal of Solids and Structures. 2015; 53: 38-47.

[12] Roux S, Rethore J, Hild F. *Digital Image Correlation and Fracture: An Advanced Technique for Estimating Stress Intensity Factors of 2D and 3D Cracks*. Journal of Physics D: Applied Physics. 2019; 42.

[13] Carrol J, Abuzaid W, Lambros J, Sehito H. *High resolution digital image correlation measurements of strain accumulation in fatigue crack growth*. International Journal of Fatigue. 2013; 57: 140-150.

[14] ASTM. *E1457-15 Standard Test Method for Measurement of Creep Crack Growth Times in Metals*. ASTM. 2015. [Online]. Available: https://www.astm.org/Standards/E1457.htm.

[15] Johnson H. *Calibrating the Electric Potential Method for Studying Slow Crack Growth*. Materials Research & Standards. 1965; 5: 442–445.

[16] Tarnowski K, Nikbin K, Dean D, Davies C. *A Unified Potential Drop Calibration Function for Common Crack Growth Specimens.* Experimental Mechanics 2018; 58: 1003-1013.

[17] Schwalbe K, Hellmann D. *Application of the Electrical Potential Method to Crack Length Measurements Using Johnson's Formula*. Journal of Testing and Evaluation. 1981; 9: 218-220.

[18] Liu M, Gan Y, Hanaor D, Liu B, Chen C. *An improved semi-analytical solution for stress at round-tip notches*. Engineering fracture mechanics. 2015; 149: 134-143.

[19] ASTM. *B670-07 Standard Specification for Precipitation-hardening Nickel Alloy (UNS N07718) Plate, Sheet, and Strip for High-Temperature Service*. ASTM. 2018. [Online]. Available: https://www.astm.org/cgi-bin/resolve.cgi?B670. [Accessed 17 May 2021].

[20] ASTM. *E8/E8M-16 Standard Test Methods for Tension Testing of Metallic Materials*. ASTM. 2018. [Online]. Available: https://www.astm.org/Standards/E8. [Accessed 1 February 2019].

[21] Paris P, Erdogan F. *A Critical Analysis of Crack Propogation Laws*. Journal of Basic Engineering. 1963; 85: 528-533.

[22] Budynas R, Nisbett J, Shigley J. *Shigley's mechanical engineering design*. New York: McGraw-Hill; 2011.

[23] Suresh S. *Fatigue of Materials.* University of Cambridge; 1998.

[24] Mercer C, Soboyejo A, Soboyejo W. *Micromechanisms of fatigue crack growth in a forged Inconel 718 nickel-based superalloy*. Materials Science and Engineering: A. 1999; 270: 308-322.

[25] Ronald J. *Fractography of Metals and Plastics.* Plastics Failure Analysis and Prevention. 2001; 127-134.

# Supplemental Material:

The following plots are supplemental data that we used to validate the claims made in this paper. Fig. S1 is the experimental stress strain curve obtained from the IN718 samples used in the experiment. This showed good accordance to the yield and ultimate strengths to what is published in literature.

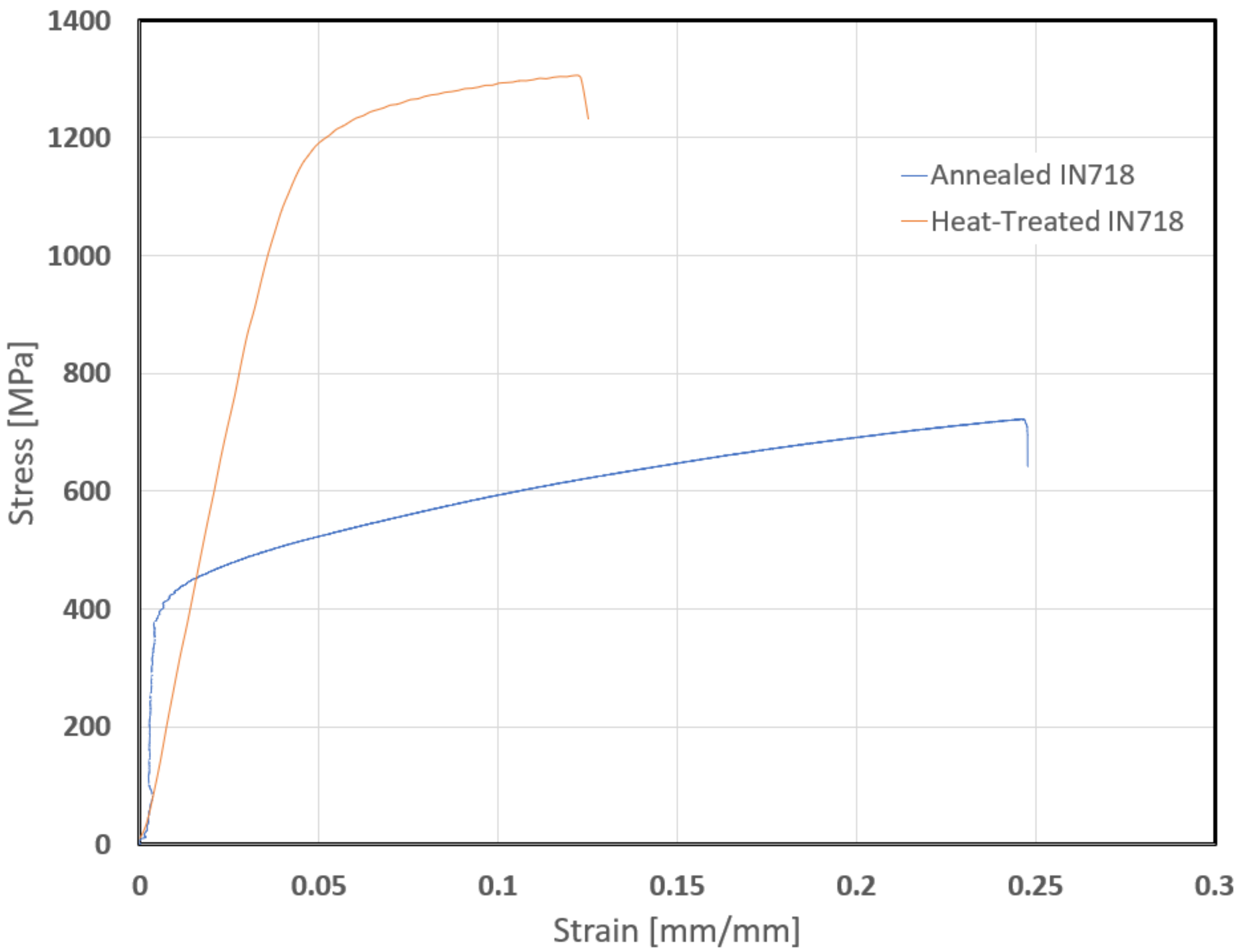

*Fig. S1 - Experimental Stress Strain Curve for Plate Inconel 718.*

Fig. S2 is a histogram of voltage measurements made over an hour with no loading. This was done to assure that the 10-amp DC signal was not influencing the data obtained using DCPD.

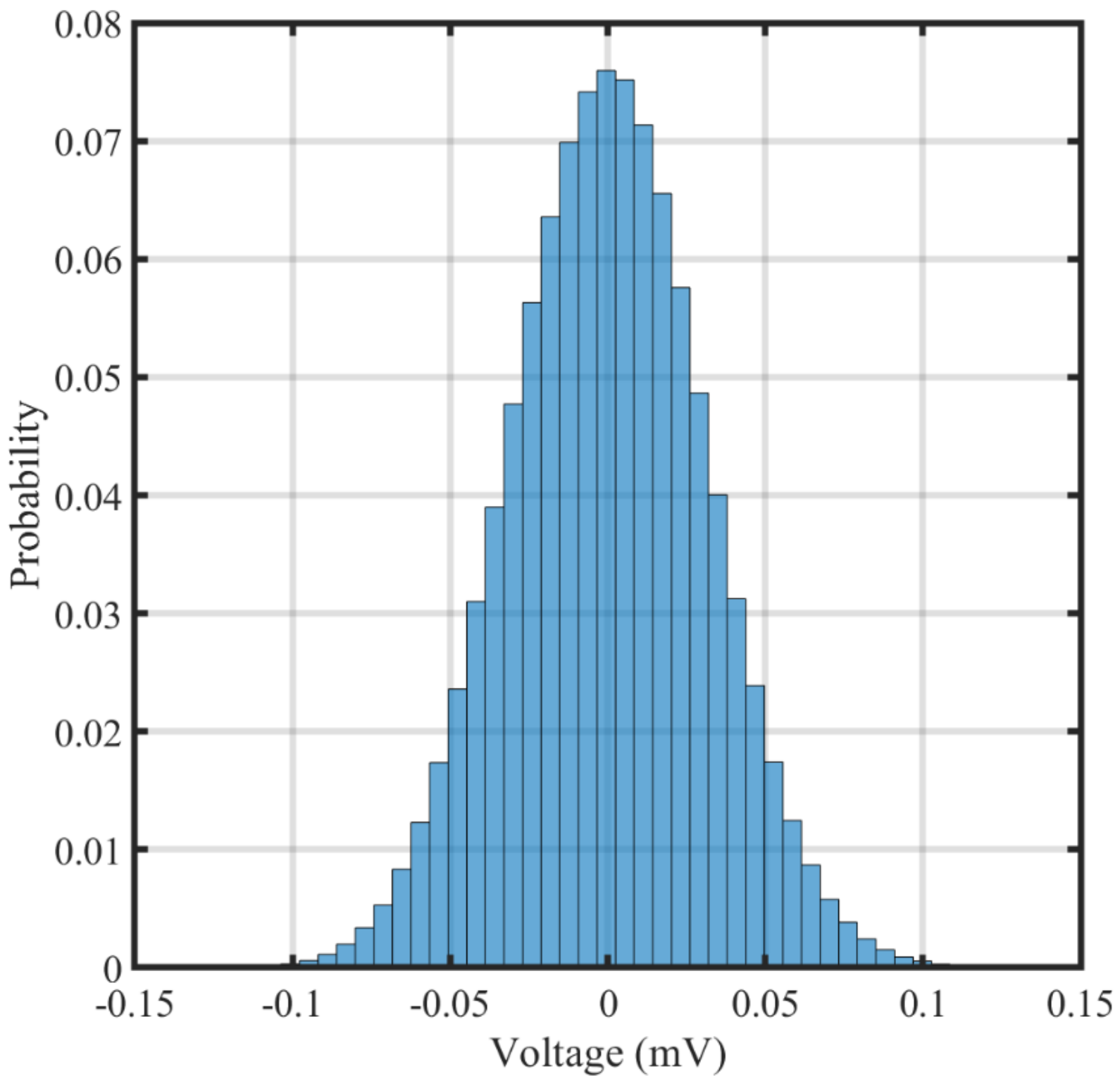

*Fig. S2 - Noise plot of a specimen under zero load to show that the electrical noise is a Gaussian distribution*

Fig. S4 through Fig. S5 are the individual distributions of crack events for each max loading.

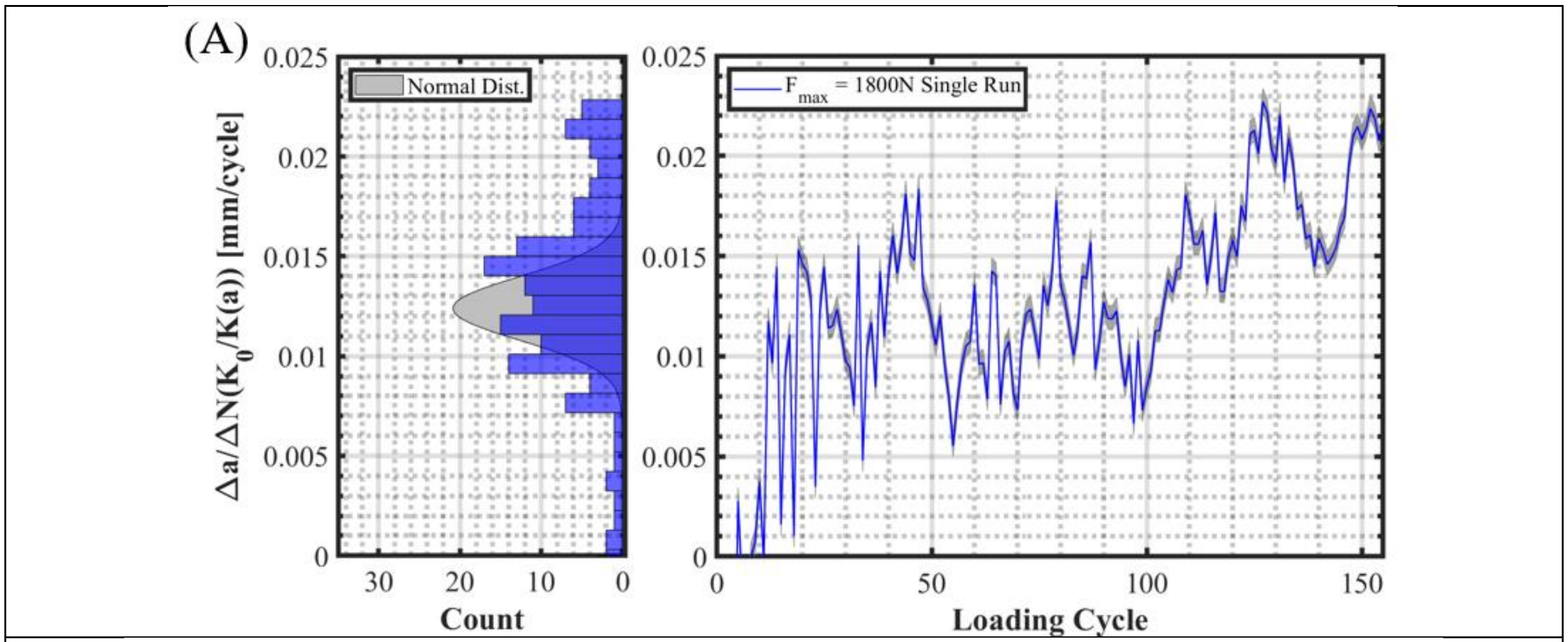

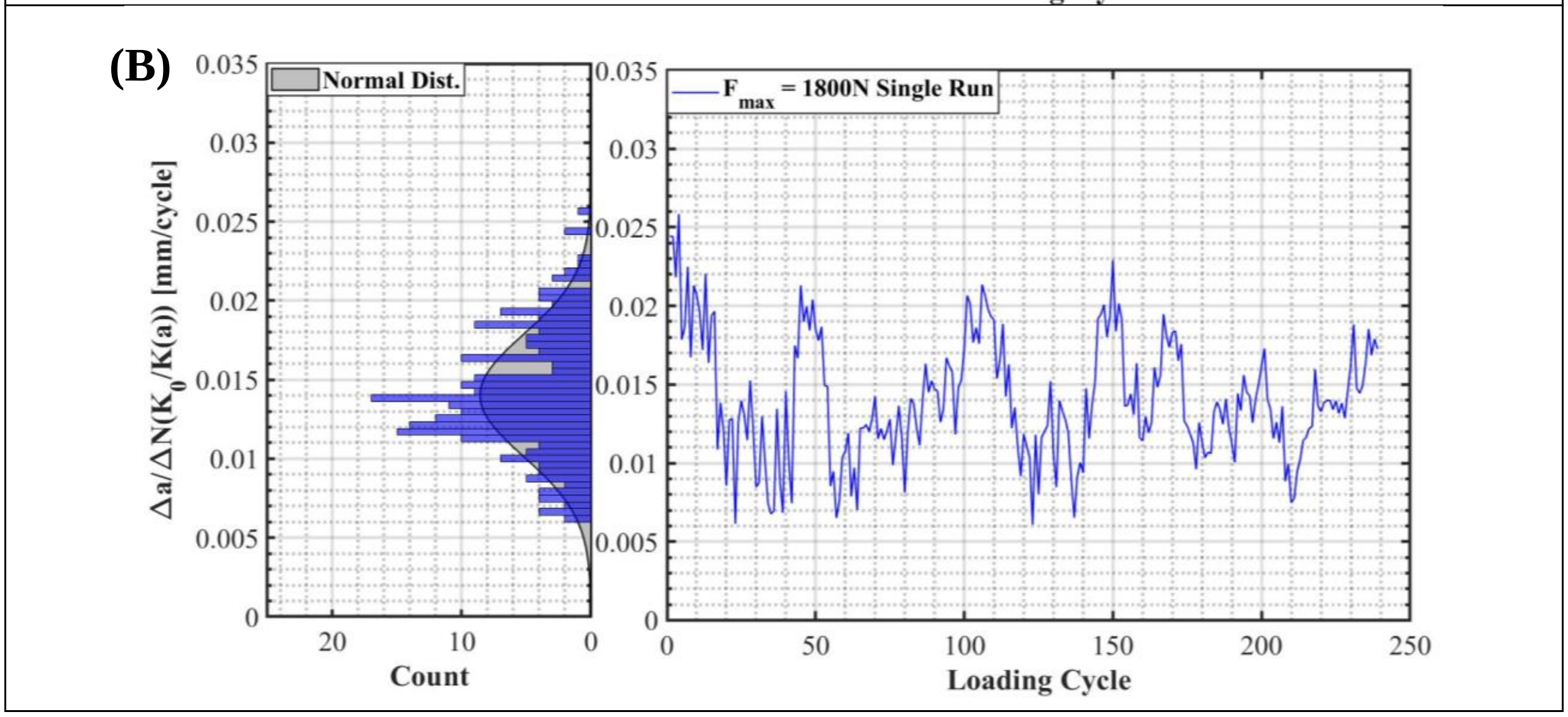

*Fig. S3(A) - $F_{max}$=1800N plot of crack jump distribution for a single as received annealed sample. Fig. S3(B) - $F_{max}$=1800N plot of crack jump distribution for a single as heat-treated sample.*

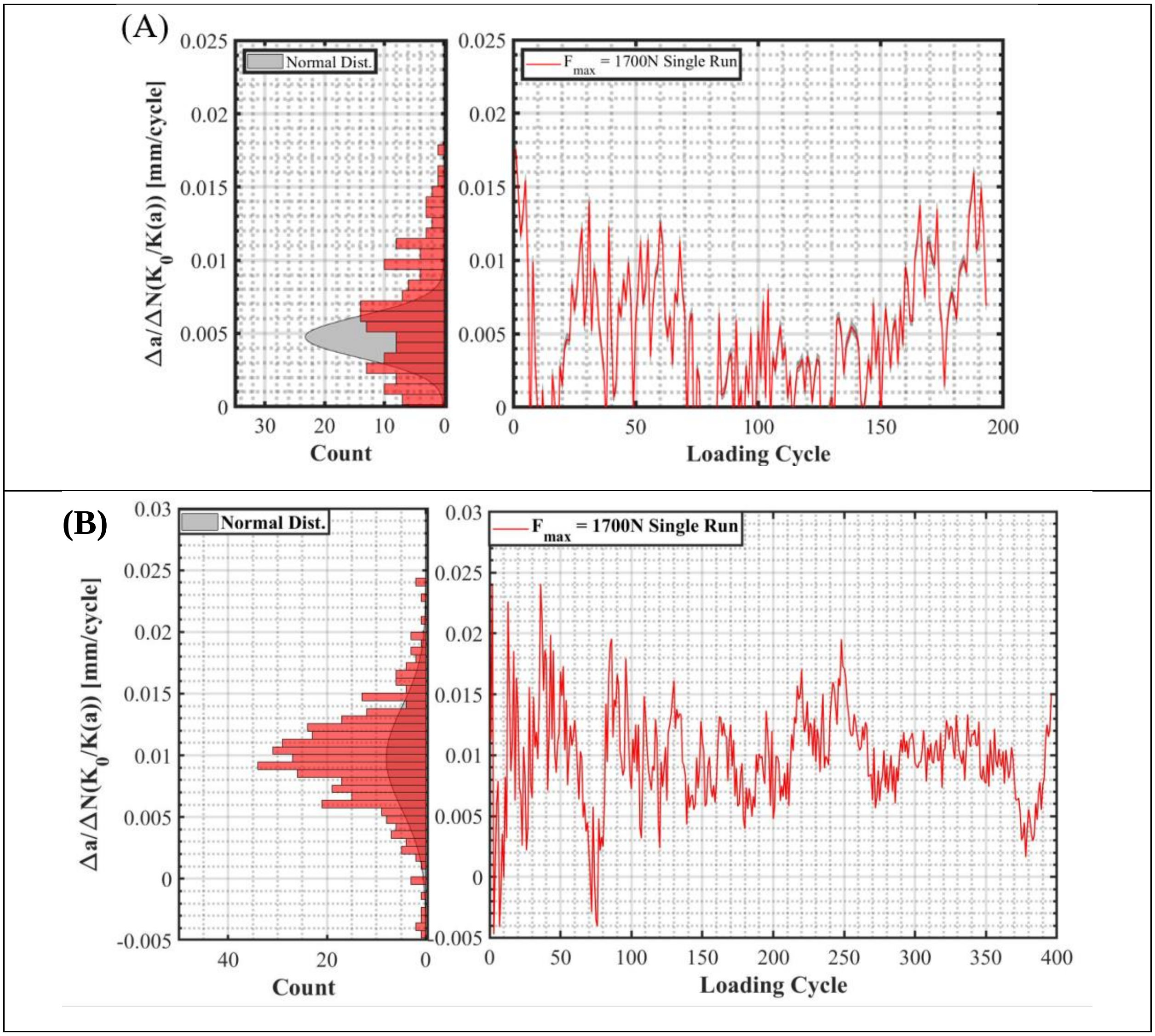

*Fig. S4(A) - $F_{max}$=1700N plot of crack jump distribution for a single as received annealed sample. Fig. S4(B) - $F_{max}$=1700N plot of crack jump distribution for a single as heat-treated sample.*

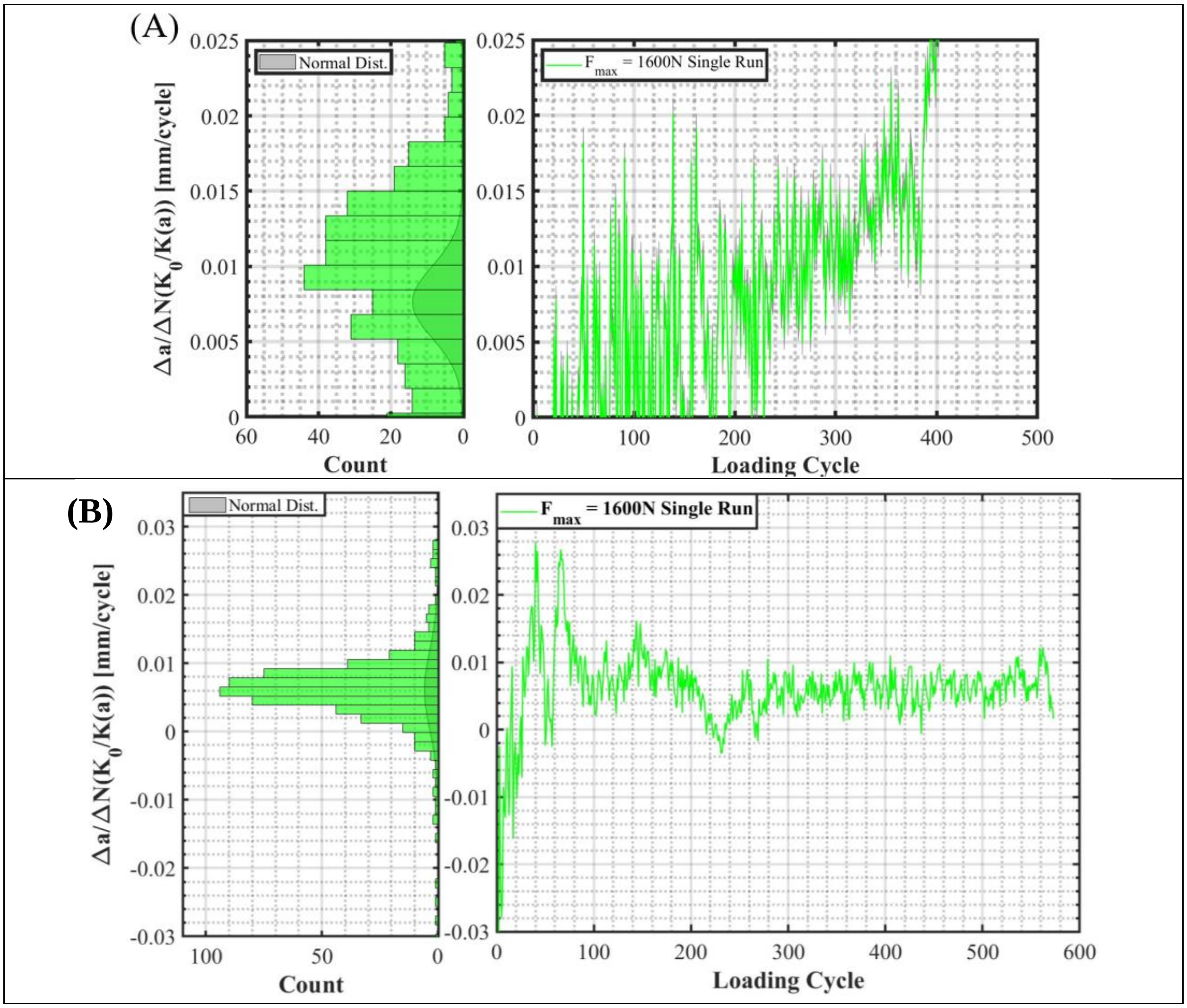

*Fig. S5(A) - $F_{max}$=1600N plot of crack jump distribution for a single as received annealed sample. Fig. S5(B) - $F_{max}$=1600N plot of crack jump distribution for a single as heat-treated sample.*

Fig. 6 is the da versus dN for the annealed specimen.

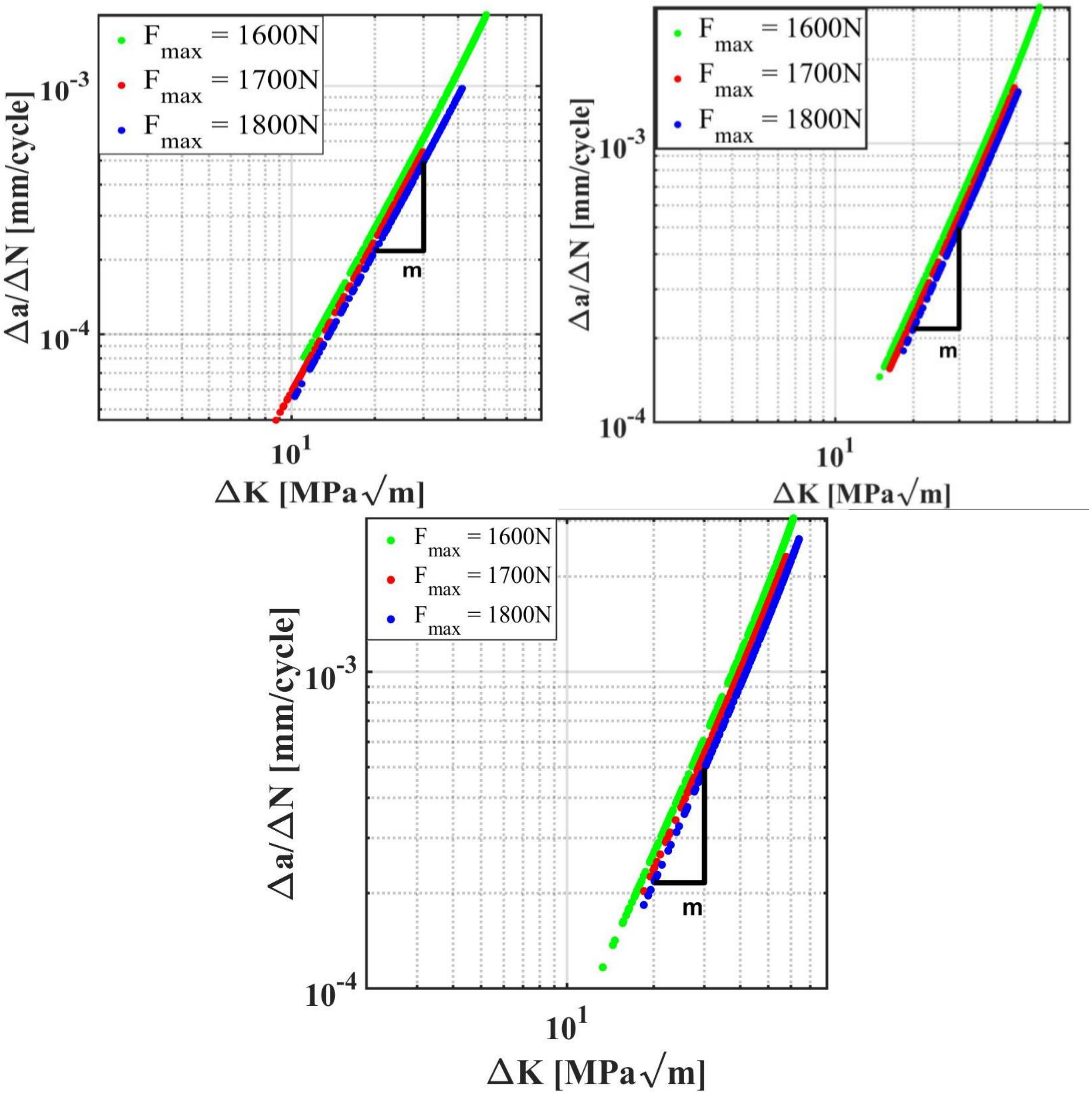

*Fig. S6(A) - Plot of Δa/ΔN versus the stress intensity factor* difference *for multiple loading cycles at RT for the annealed samples. Fig. S6(B) - Plot of Δa/ΔN versus the stress intensity factor* difference *for multiple loading cycles at 185°C for the annealed samples. Fig. S6(C) - Plot of Δa/ΔN versus the* stress intensity factor difference *for multiple loading cycles at 285°C for the annealed samples. The data from all experimental datasets demonstrate Paris' Law regime.*